# Evolutionary sparse data-driven discovery of complex multibody system dynamics

Ehsan Askari[1,2,*], Guillaume Crevecoeur[1,2]

[1]*Department of Electromechanical, Systems and Metal Engineering, Ghent University, B-9052 Zwijnaarde, Belgium*

[2]*EEDT Decision & Control, Flanders Make, Lommel, Belgium*

## Abstract

The value of unknown parameters of multibody systems is crucial for prediction, monitoring, and control, sometimes estimated using a biased physics-based model leading to incorrect outcomes. Discovering motion equations of multibody systems from time-series data is challenging as they consist of complex rational functions, constants as function arguments, and diverse function terms, which are not trivial to guess. This study aims at developing an evolutionary symbolic sparse regression approach for the system identification of multibody systems. The procedure discovers equations of motion and system parameters appearing as either constant values in function arguments or coefficients of function expressions. A genetic programming algorithm is written to generate symbolic function expressions, in which a hard-thresholding regression method is embedded. In an evolutionary manner, the complex functional forms, constant arguments, and unknown coefficients are identified to eventually discover the governing equation of a given system. A fitness measure is presented to promote parsimony in distilled equations and reduction in fit-to-data error. Hybrid discrete-continuous dynamical systems are also investigated, for which an approach is suggested to determine both mode number and system submodels. The performance and efficiency of the suggested evolutionary symbolic sparse regression methodology are evaluated in a simulation environment. The capability of the developed approach is also demonstrated by studying several multibody systems. The procedure is efficient and gives the possibility to estimate system parameters and distill respective governing equations. This technique reduces the risk that the function dictionary does not cover all functionality required to unravel hidden physical laws and the need for prior knowledge of the mechanism of interest.

---

[*] Corresponding author

E-mail address: ehsanaskary@gmail.com (E. Askari)

# 1. Introduction

Mathematical equations governing the dynamics of systems are of paramount importance in both science and technology, linking the system input to the output. Having a mathematical model of a system, one can simulate, predict, control, and diagnose the system. One of the approaches to constructing a mathematical formulation is well-known as system identification which estimates the model based on the observed time-series data collected from a system for which either nothing or limited prior knowledge is available [1,2]. Zadeh [3] suggested a definition for system identification "System identification is the determination on the basis of observations of input and output, of a system within a specified class of systems to which the system under test is equivalent." Depending on the application, one might find a trade-off between the simplicity and accuracy of the identified model [3, 4]. For example, in control engineering, one might be interested in just a simple form of an engineeringly acceptable model to give an approximation of the height level of water in a fluid-storage container, while in precision tool engineering, one looks for the most accurate model that matches the machine to a very high level.

The field of multibody system dynamics focuses on the study of the dynamic behavior of bodies that are connected by joints and force elements such as spring, damper, and actuators, restricting their relative motion [5, 6]. The application of multibody systems is very wide in the industry from automobiles, spaceships, aircraft, and assembly lines to human body motion. Loads generated by moving parts of a multibody system are sometimes difficult to measure. Control of their motion, especially the precise movement of robots and tools, is another challenge with such systems. In addition, condition monitoring of such systems plays an essential role in the improvement of their longevity and prevention of their catastrophic failures. In the process of product design, one might also need to gain an insight into the way multiple moving parts of a multibody system interact with each other and if they provide the designer with the reachability and performance needed [7]. All these challenges can be addressed to some extent, provided that one discovers the mathematical model governing the respective system.

On top of that, the value of system parameters used in the theoretical models of multibody systems plays an essential role in accurately predicting the response of a physical system [8]. Parameter estimations of multibody systems also are crucial for prediction, monitoring, and control. The common practice is to estimate system parameters using a physics-based model that is derived from the physics of the problem like in classic mechanics, Fig. 1. In classic mechanics, governing equations of a system are derived on the basis of, for example, the mass and momentum conservation laws, and principal thermodynamics laws. It is well-known that developing such constitutive models requires a good knowledge of the system and its environment [9]. The complexity of machines and multi-physics phenomena involved, environmental conditions, and a lack of information on how system parameters vary over time hinder the construction of efficient physics-based models or at least make their development very difficult [10].

An alternative solution is to go for data-driven models extracted based on data science and data collected from a system of interest. In data science, there is a great possibility to integrate statistical learning concepts with classical approaches in applied mechanics and mathematics to discover sophisticated and accurate models of complex dynamical systems directly from data [11, 12], as is illustrated in Fig. 1. Such data-driven models are acquired using Pareto front, sparse regression methods, equation-free modelling, empirical dynamic modelling, modelling emergent behavior, and automated inference of dynamics [11, 13-16]. These data-driven model discovery approaches have successfully been used in the research field of fluid mechanics, material engineering, and dynamical system to obtain governing equations [17, 18]. These open the possibility to discover the governing equations of a given system just from time-series datasets in the spatial domain, thanks to the advancement of data science.

When considering multibody systems, available data-driven approaches commonly suffer from a need for prior knowledge and expert intuition of the system, and a lack of function diversity especially when rational functions are needed [19-21]. Previously developed approaches do well when systems under consideration have equations consisting of polynomial functions, trigonometric functions with integer constant-valued arguments, among others. However, there are multibody systems that are governed by models in which constant-valued arguments belong to the set of real numbers, e.g., 2.35 in sin (2.35 *t*), which cannot be addressed by available approaches. It is worth mentioning that Schmidt and Lipson [14] developed a methodology based on the genetic algorithm to discover natural laws from experimental data, in which constant values were estimated. However, their interesting method was not able to generate random function sets to cover the function diversity required to cover different types of physical problems. Systems with rational functional forms were also investigated by Mangan et al. [19] where they recast the dynamics model with rational functions in an implicit form and obtained the solutions in null space. They used their method to discover three canonical biological models with simple polynomial terms in both nominator and denominator.

In the era of the fourth industrial revolution (Industry 4.0) that is the industry trend and activity toward automation and data exchange in the industry including manufacturing technologies and processes, a huge amount of data is available due to the affordable cost of sensors and increasing smart machines and systems [22]. Therefore, the present study aims at developing a data-driven methodology to discover the motion equations of multibody systems from their time-series dataset. An evolutionary symbolic spare regression approach is suggested in which symbolic function expressions are randomly constructed. This study employs a ridge-regression method to acquire the unknown coefficients in generated governing equations. The values of constant arguments of candidate terms are also estimated based on a specific mutation technique presented in this study. In addition, a fitness measure is suggested, which promotes parsimonious equations of motion and reduces not only the complexity of individuals and their terms but also fit-to-data error.

There are multibody systems that do not operate in just one mode and switch between several dynamical regimes each of which is governed by a different model [23, 24]. To name a few, one can mention frictional sliding, contact/impact systems, biomedical mechanisms like the knee joint with nonlinear ligaments, mechanical systems with sudden emerging defects and loadings, among others [24, 25]. These systems exhibit both continuous and discrete behavior. The state of such a system can be defined by continuous variables, while modes are valued discretely. The dynamic system operates continuously in one mode as long as some constraints are held. When an event occurs, the transition can discretely take place. System identification of such dynamical mechanisms is not trivial and requires a high knowledge of the system to derive physics-based models. One does know neither the model nor the transition map and the sequence of dynamic modes. Therefore, prior knowledge of the hybrid dynamical system is required along with expert intuition to eventually assign submodels to each dynamic mode. Mangan et al. [26] employed sparse regression methods to identify hybrid dynamical systems and they used their developed approach to obtain a mass-spring hopping model and an infectious disease one, both of which can be considered simple. The present article develops a methodology to study hybrid dynamical multibody systems. Time-series data are clustered, and, subsequently, the algorithm identifies the number of dynamic modes, submodels, and switching sequences. A sliding mass subjected to friction is investigated to evaluate the performance of the suggested method.

The contribution of this study to the field of multibody system dynamics can be listed as follows: (*i*) a dedicated data-driven approach to discover the motion equations of multibody systems is developed; (*ii*) a customized methodology is suggested for hybrid dynamical system to determine number of modes, submodels, switching sequences, and transition boundary; (*iii*) the procedure mitigates the risk that the function dictionary does not cover all functionality required for the model discovery as well as the need for prior knowledge of multibody systems; (*iv*) the developed methodology does discover complex rational functional forms that are commonly encountered in the motion equations of multibody systems; (*v*) a methodology is presented that can discover the model even with small size data and is robust against noise; (*vi*) the programming algorithm developed is completely independent from commercial software tools and available codes, and can be used anywhere the data can be collected

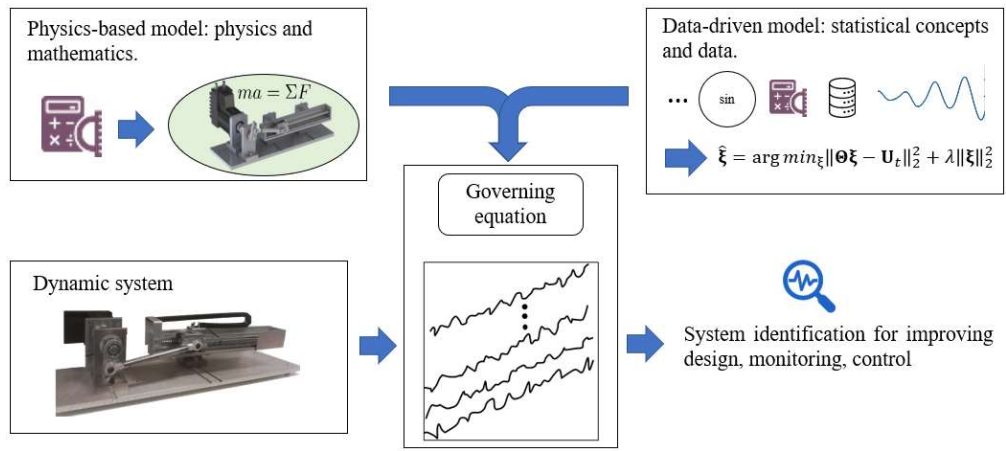

Fig. 1. System identification applied to a dynamic system.

## 2. Multibody system dynamics

Multibody system dynamics is the study of the dynamic behavior of bodies interconnected by kinematical joints and force elements such as spring, damper, and actuator [5, 6], as is demonstrated in Fig. 2. The motion of a multibody system is described by the equations of motion obtained from either Newton-Euler equations or Lagrange's equations to which constraint conditions are added [27]. With the augmentation of constraint equations to the equations of motion using the Lagrange multiplier method, the motion of such a system can be expressed by the following differential algebraic equations [5, 28, 29]

$$\mathbf{M\ddot{q}} + \mathbf{C}_q^T \boldsymbol{\lambda} = \mathbf{F}, \quad \text{s.t.} \quad \mathbf{C}(\mathbf{q}, t) = \mathbf{0} \tag{1}$$

where **M** and **q** are the mass matrix and generalized coordinates of the system including both translation and rotation coordinates, respectively. The equations of motion written in this format uses redundant coordinates because the number of coordinates used in Eq. (1) is commonly more than the degrees of freedom of the dynamic system. The load vector is depicted by **F** that conveys external forces along with Coriolis and centrifugal terms acting on the system bodies. **λ** are Lagrange multipliers and **C** represents the holonomic algebraic constraints while its derivatives with respect to the coordinates is designated by $\mathbf{C}_q$, which is called the constraint Jacobian. One may transform Eq. (1) to ordinary differential equations such that the ordinary integration methods are employed to integrate them over time. Differentiating the constraint equations twice, the equations of motion of multibody system dynamics can be cast as follows

$$\begin{bmatrix} \mathbf{M} & \mathbf{C}_q^T \\ \mathbf{C}_q & \mathbf{0} \end{bmatrix} \begin{bmatrix} \ddot{\mathbf{q}} \\ \boldsymbol{\lambda} \end{bmatrix} = \begin{bmatrix} \mathbf{F} \\ \boldsymbol{\gamma} \end{bmatrix}, \quad \boldsymbol{\gamma} = -(\mathbf{C}_q \dot{\mathbf{q}})_q - 2\mathbf{C}_{qt} - \mathbf{C}_{tt} \tag{2}$$

This equation can be integrated over time using a standard numerical integration procedure. However, the equation (2) cannot ensure the satisfactory incorporation of constraint equations appearing in Eq. (1) as here are the acceleration constraints considered. Therefore, numerical integration can accompany with errors and eventually diverge. To mitigate this issue, employing a stabilization method is suggested [28]. As redundant coordinates are used to formulate the system motion, the coordinates are dependent on each other, which is why the constraint conditions should simultaneously be solved. In comparison, solving the equations of motion with minimal coordinates that are the same as degrees of freedom can be more economical and easier due to the absence of constraint conditions and the use of the standard time integration methods without any concern about the numerical stability due to constraints. However, the transformation of the redundant coordinates to the minimal ones is computationally expensive [30, 31].

In this study, let's consider one can write equations of motion using minimal coordinates as follows [14, 23]

$$\dot{\mathbf{z}}(t) = \mathbf{f}(\mathbf{z}(t)) \tag{3}$$

in which $\mathbf{z} = [\mathbf{q} \quad \dot{\mathbf{q}}]^T$, $\mathbf{z}(t) \in \mathbb{R}^n$, represents the system state at time $t$ while $\mathbf{f}(\mathbf{z}(t))$ is the nonlinear physics-based function expressing the motion of the multibody system. In this paper, a system identification is presented to achieve two goals: (*i*) to discover nonlinear governing equations, Eq. (3); and (*ii*) to estimate parameters, directly from time-series datasets collected from a given multibody system.

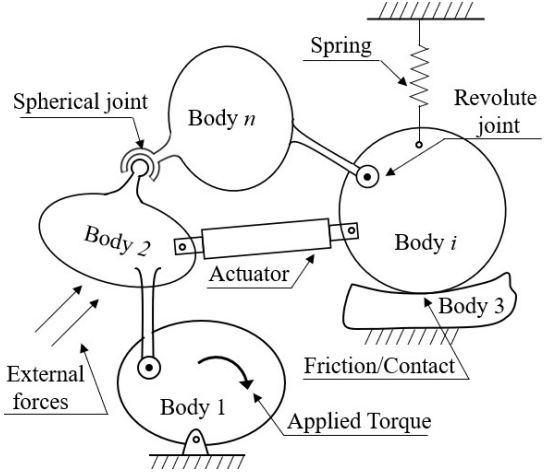

Fig. 2. A schematic representation of a multibody system.

## 3. Time-series datasets collected from a given system

Several sets of measurements of a given multibody system at time points $[t_1, \cdots, t_m]$, $\mathbf{z}(t_i) \in \mathbb{R}^n$, from multiple initial conditions, which may come from a different set of time points, are collected, and concatenated together being presented by $\mathbf{Z} = [\mathbf{z}(t_1), \mathbf{z}(t_2), \cdots, \mathbf{z}(t_Q)]$ [26]. The dataset is divided into two sets, namely training set, $\mathbf{Z}_T \in \mathbb{R}^{m \times n}$, and validation set, $\mathbf{Z}_V \in \mathbb{R}^{v \times n}$, in which *m* and *v* are the number of training samples and validation samples, respectively, and their summation is equal to *Q*. Such a set of time-series data can be either full, consisting of both system states and their time-derivatives or incomplete. When the input is not complete, one needs to complete it by either differentiating or integrating the data with respect to or over time. If the data is noise-free, time derivation is very straightforward using the finite difference method to do the job [32]. On the other hand, the data corrupted with noise is troublesome, which is later discussed in Section 4.1. The system identification procedure developed in this study begins with data collection that is fed to an evolutionary symbolic sparse regression module, shown in Fig. 3, where the analysis is carried out on the data and eventually the discovered parsimonious model is given as the output.

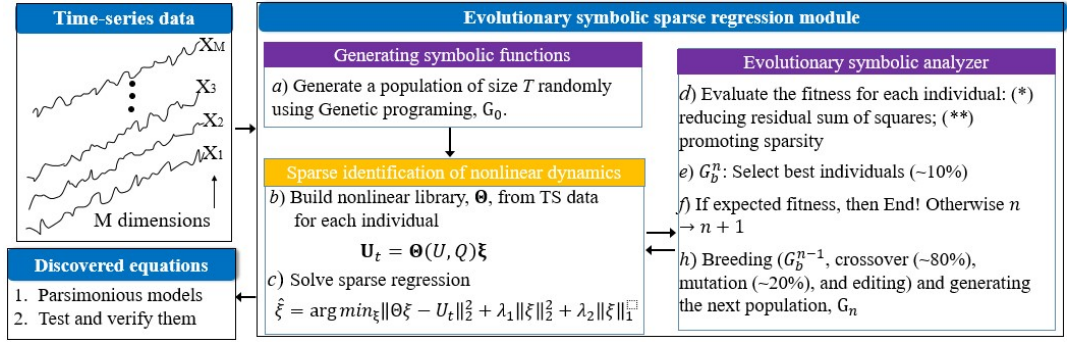

Fig. 3. The workflow of the methodology presented in this study.

# 4. Artificial intelligence: evolutionary symbolic sparse regression module

This study aims at developing an artificial intelligence (AI) system to interpret time-series datasets, to analyze and, subsequently, learn from the input data according to an embedded optimization functional towards maximizing the chance to achieve the predefined goals that are to identify the respective physics-based model and to estimate system parameters, followed by inferring actions. The evolutionary symbolic sparse regression module, demonstrated in Fig. 4, is the core of the artificial intelligence system, which consists of data treatment, genetic programing, sparse regression, and inferring actions. The aims this AI system follows are (*i*) discovering the physics-based model and (*ii*) estimating parameters of a given multibody system.

### 4.1. Data treating – smoothing, incomplete and noisy data

In practice, there is a situation in which state variables are available as time-series datasets, and one needs to estimate their derivatives employing numerical approaches. Finite difference methodology (FDM) is the widely used method that unfortunately does not do well when the data is corrupted with noise. It is important to note that denoising the data either before or after the differentiation using the FDM does not lead to satisfactory derivatives. The other type of methods, including the Savitzky-Golay filter, gives the possibility to fit a local model of the data, e.g., a sliding polynomial with a low-degree polynomial, via linear regression [33]. Approaches like Tikhonov regularization and total variation regularization directly regularize the differentiation process [32, 34]. In this study, three procedures including FDM, the Savitzky-Golay filter, and total variation regularization are employed for the time-differentiation of noisy time-series (TS) datasets, which gives the possibility to do an efficiency comparison among them. In addition, the data collected from a given machine is corrupted with noise and, in turn, need to get smoothed somewhat before being fed to the artificial intelligence system [35]. The methods like the Savitzky-Golay filter, Tikhonov regularization, and total variation regularization can be utilized to smooth TS data.

## 4.2. Genetic programming

The genetic programing is an extension of genetic algorithm, which was initially suggested to answer "how the computers learn to solve problems without being explicitly programmed?" [36]. The genetic programing was demonstrated that can successfully get the computer programmed by means of natural selection [37-39]. The genetic programing (GP) paradigm is applicable to a broad range of problems in optimization, machine design, control engineering, and system modeling, to name a few, as it gives the possibility to find an optimal plan and algorithm for problems [37]. In this article, the GP is to seek a set of candidate functions that form the physics-based model of a given multibody system. The optimization problem is that the discovered model represents time-series data collected from a given mechanism and such a model is not complex [14, 20, 21]. The workflow of the suggested method is presented in Fig. 3, which consists of data collection and the AI module in which the GP and sparse regression are integrated. The flowchart of the algorithm associated with the evolutionary symbolic sparse regression module in detail can also be observed in Fig. 4.

Figure 4 shows that the process begins with treating the data either time-differentiation or smoothing before generating a population randomly. Population consists of *N* individuals each of which includes a number of candidate terms (minimum one). The GP algorithm creates each function using a tree structure and based on two predefined sets: (*i*) terminal set; and (*ii*) function set [20, 37]. The terminal set includes constants, and both system and input variables, Eq. (4), while the function set consists of basic mathematical operations and a subset of the elementary functions, Eq. (5). The members of the latter are defined such that they can construct potential solutions to a given dynamics problem along with the terminal set. Although the function set can include user-defined functions according to prior knowledge of the system, the developed algorithm does not consider such expert-defined functions to prevent any bias in the solution. However, the reader can later see that the algorithm does take advantage of the knowledge produced by itself for future use. In addition, there are systems whose equations of motion include integrands that need to be integrated with respect to either time or system state, among others. The suggested algorithm gives this possibility to include the integration operator in the function set such that randomly generated integrands are integrated over time.

$$\{t, c, 1, z_1, z_2, \ldots, z_n, \ldots, u_1, u_2, \cdots\} \tag{4}$$

$$\{+, \times, \div, \sin, \text{sqr}, \text{abs}, \exp, \text{sgn}, \text{pwr}, \text{int}, \log, \arcsin \ldots\} \tag{5}$$

where 'sqr' is the square root $\sqrt{\phantom{x}}$, 'abs' stands for the absolute function $|\phantom{x}|$ while 'pwr' and 'int' are, respectively, exponentiation function with real exponents and integration operator. An example of an individual generated by the GP algorithm is presented in Eq. (6) in which 'hvs' depicts Heaviside function and 'sgn' stands for Signum function. It is worth noting that each constant presented by symbol *c* appearing in the functions' arguments is assigned a random value, which can be confined to be between minimum and maximum values set by the user. In operations inspired by biological evolution to generate new populations, previously assigned constants do not vary unless they are chosen to undergo mutation and crossover actions. In the following, main steps introduced in the flowchart, Fig. 4, are discussed.

$$\frac{z_4}{\exp(z_3 z_1)}, \text{hvs}|z_4|, \exp z_3, \frac{z_2^2 z_4}{\sin c}, t, \text{sgn}(c)(z_3 z_2 + z_3^2), 1, \sin(c + ct), |z_1|, z_3, \sqrt{t}, z_2, |z_1 \sin z_1^2|, \frac{z_2 \sin z_3}{t}, \frac{c + \sin z_1}{\sqrt{t + z_2 z_3}} \quad (6)$$

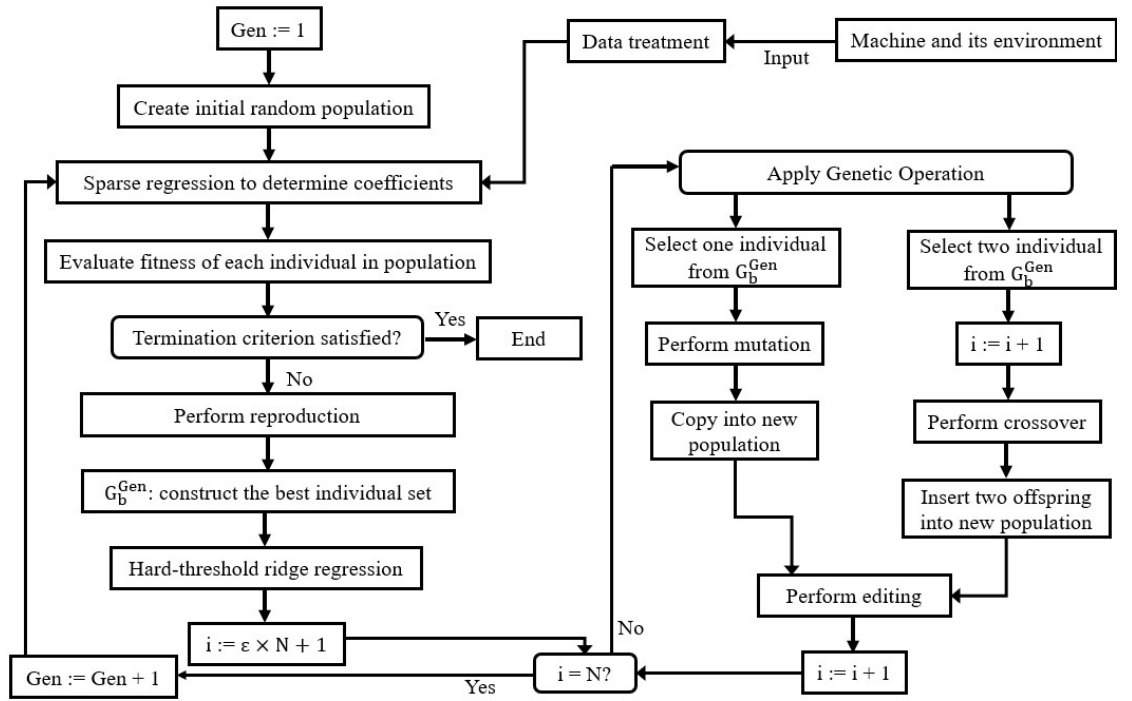

Fig. 4. Flowchart for the evolutionary symbolic sparse regression module.

### 4.2.1 Breeding the next generation

Operations used to generate new population while modifying the structures in genetic programing are categorized into two groups: (*a*) primary operations and (*b*) secondary operations [37, 38]. Reproduction and crossover operations belong to the former category, whilst mutation and editing can be named as two of the secondary operations. In the following, a general description of them is given while their specific characteristics used in the GP algorithm developed in this study are detailed.

### (a). Primary operations: reproduction and crossover

The reproduction operation does select a single expression from the population based on a fitness-based selection method and just copy it to the next generation (population). In this study, a set of the individuals is constructed by sorting the population based on the better fitness, which means the individual with the best fitness places as the first expression in the list. A percentage of the members of that set is selected, that is, $\varepsilon_1$% commonly 10%, and placed in a new set, $G_b^n$, called the set of the best individuals. The selection method used in this algorithm to do reproduction is called the tournament selection that, randomly, chooses a group of members (two or three) belonging to the current population while excluding members of $G_b$ and the one with the better fitness is the final expression, unchanged, to be copied to the new generation (population). The reproduction operation forms $\varepsilon_2$%, commonly 10%, of the new population in the developed algorithm, which is added to $G_b^n$ including $\varepsilon$% of the population.

The crossover operation introduces new variants and variation in the population whilst randomly chooses two parents and produces two new offspring possessing parts from each of those parental expressions, as is demonstrated in Fig. 5 [37, 40]. Each of parental expressions are randomly chosen using the tournament selection method. It is worth noting that each individual includes several candidate terms (functions) and one function is randomly selected from each of two randomly chosen individuals provided that both chosen functions are not operands simultaneously. It is noted that two parents are typically of different sizes and after crossover action, the size of resultant terms may become very large. A strategy used in this article is to cancel crossover outputs with the size more than 15, counting terms from function and terminal sets in row, and redo the process. Moreover, the crossover operation must lead to meaningful functions. For example, replacing a subtree from a point (node) that contains mathematical operation with a member of the terminal set, such as $\boxed{\times}\, z_1 t \to \boxed{z_2}\, z_1 t$, does not produce a meaningful output. The algorithm uses a uniform probability distribution to choose randomly points or nodes on the trees of parental expression to prevent any bias in operation.

In the string format presentation of each chosen offspring parent, the type of characters is detected whether are mathematical operations or operands and, subsequently, restored in an array with the same sequence they appear in the string format of the corresponding tree (function). One random node in each parental tree is selected, that is, the crossover node for that parent. The crossover fragment is formed from the crossover point and includes the whole subtree below the crossover node going towards the far distance from the original node of the tree. Crossover fragments of two parents are exchanged and placed at the corresponding crossover points [37]. The algorithm also gives the possibility to choose just one terminal or even the whole tree as the crossover fragment of a parent. The procedure done by the crossover operation can be observed in Fig. 5. As an outcome of the crossover process, it is observed that a functional operator such as sine function repeats itself more than two times in row, which is deemed to be incompetent. The developed algorithm kills the repetition of a function that occurs more than two times in row, for example $\sin \sin \sin \sin z_1 \to \sin \sin z_1$. The crossover operation is also implemented to constants appearing in function arguments. The structure of the candidate terms is kept and just a randomly selected constant in a randomly chosen function is replaced by one of those candidate terms belonging to the set of best individuals. As an outcome of this process, just one new individual is generated. In addition to the logic behind the crossover action, this process leads to reduce complexity of function terms by decreasing the number of constants with different magnitudes. This outcome is very beneficial in the case of rational functions in particular, which is commonly seen when working with multibody system dynamics. Within the structure of the algorithm, the crossover operation allocated to constants is placed inside the subroutine of mutation.

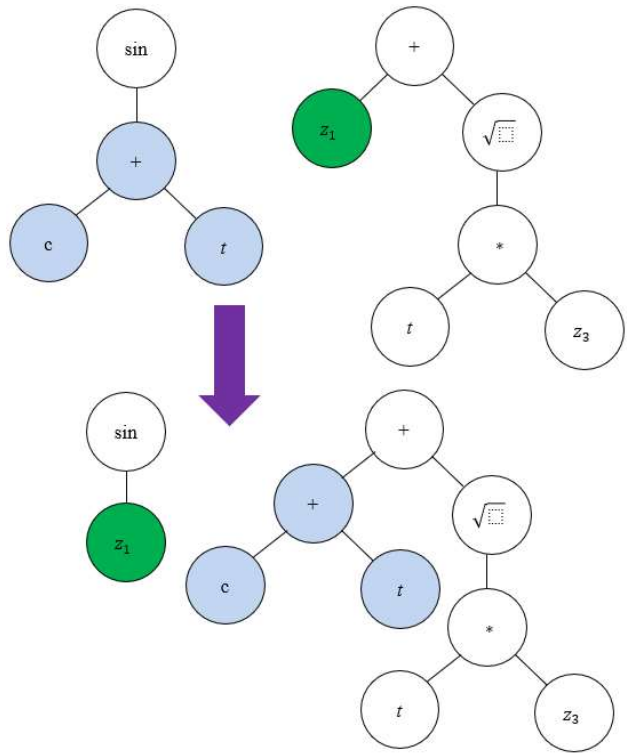

Fig. 5. Crossover operation of two candidate functions $\sin(ct)$, written as $\sin * ct$ by the algorithm, and $z_1 + \sqrt{tz_3}$, in the algorithm $+z_1 \text{sqr} * tz_3$ (sqr: square root), as parental expressions, which produce two following offspring trees $\sin(z_1)$ and $ct + \sqrt{tz_3}$.

**(b). Secondary operations: editing and mutation**

In the process of population generation, trees are created that can be simplified, removed, or modified. The editing operation is dedicated to such situations. This operation does take required actions when any of the following scenarios are observed.

- Remove repeated candidate terms in an individual.
- There are, for example, functions such as square root, i.e., $\sqrt{\Xi(z, t, \dots)}$, mapping the set of nonnegative real numbers onto itself. If such a function term gets a negative value inside the radical sign due to the random nature of the genetic programing and symbolic regression method, the editing module removes the tree.

- Simplification of candidate terms such as "$\sin c$" and "$/z_3 z_3$" both of which can be converted to "1".
- Simplification of functions like signum and absolute, which show specific characteristics, for example, $t > 0 \rightarrow \text{sgn} t = 1$ and $\text{abs} t = t$. Therefore, these types of functions can also be replaced by elements belonging to the terminal set, namely 1 and *t* in the case of the above examples. They can subsequently be removed if the corresponding individual already includes those resultant elements, for example *t* and 1. This process can also be performed when the library matrix is built, and the similar vectors are detected.
- In the case that the function terms are not independent, they produce library matrix whose column rank is less than the column numbers (it is assumed that the number of rows is more than the column's). This is checked in the library matrix to remove those suggested functions that are dependent.
- The removal of rational functions generated by the algorithm, whose denominator gains zero value.

The aim of doing mutation is to randomly introduce alterations in the population structure and it is beneficial promoting diversity in a population that may intend to converge prematurely [37]. It is worth noting that the mutation is a secondary operation in the genetic programing and operates on one parental expression that is selected from $G_b$ using the tournament selection method. This operation is controlled such that the maximum depth (size) of the new subtree created during the mutation operation does not become more than the initial maximum depth size specified for parental trees or expressions. The mutation takes two following actions: (*b*) make a change in constants; (*c*) a change in subtree randomly. The latter, demonstrated in Fig. 6c., is performed by randomly replacing a randomly chosen subtree of a randomly chosen function (tree) in a randomly chosen individual belonging to the set of best individuals obtained in the previous generation with a new tree randomly generated. The mutation also operates on the constants appearing as function arguments, Fig. 6b, which is not common in practice. As the mutation is going to be done on a parental expression among the set of best individuals, slight modifications on function arguments can be helpful adjusting the discovered function term. Hence, a modification according to the number of generations is applied to a constant employing the following relationship

$$c_{new} = c_{old} + \frac{2\kappa - \vartheta}{\vartheta + 2j}, \qquad 0 \leq \kappa < \vartheta \tag{7}$$

in which *j* and $\kappa$ indicate the generation number and a randomly value between 0 and $\vartheta$ set by the user, respectively. It is worth noting that the value that a constant can gain during the process of the function generation is randomly selected in a range the user sets.

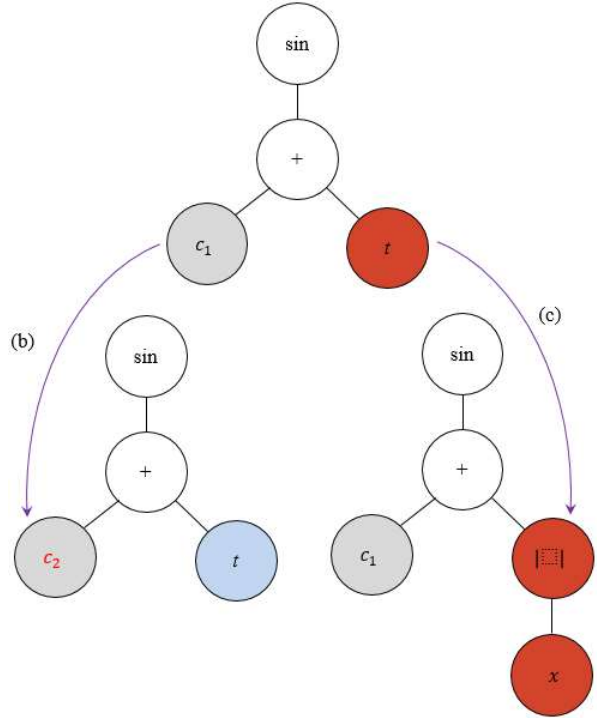

Fig. 6. Mutation operation: (*b*) a change in constants; (*c*) a change in subtree.

*4.2.2 Fitness function*

Fitness is the driving force of Darwinian natural selection and, likewise, of genetic programming [37]. Measuring fitness in our mathematical algorithm controls the application of the operations that eventually leads to the modification of the structures in the artificial population. Each individual belonging to the population is evaluated based on a fitness function defined within the algorithm. The goal is to find the individual that represents the data the best. Therefore, the residual sum of squares (RSS) is used to estimate the error between the output, $\dot{\mathbf{Z}}$, and $\mathbf{\Theta}_P^n \mathbf{\xi}$ in which the library matrix $\mathbf{\Theta}_P^n$ is constructed from the individual *P* of the population associated with the $n^{\text{th}}$ generation and the coefficients $\mathbf{\xi}$ are obtained from sparse regression (Section 4.4). On top of that, one looks for parsimonious models one of which is eventually selected as the solution. Thus, the other aim is to promote the parsimonious individuals. Achieving both goals requires a multi-objective fitness function in the GP algorithm. Such an objective functional can prevent the occurrence of bloating effect that is caused when individuals lengthen excessively [21]. In addition, filling the population with many solutions that produce low fitness, leading to the slowness of the algorithm, is modified. This also avoids overfitting as the model accuracy needs to be balanced with its complexity [41]. When there is a fitness function with two objectives, a good strategy is required to compromise between their effects or weights on the final selection.

An adaptive fitness measure is employed in this study. Two thresholds on both the number of generations, $N_{\text{threshold}}$, and fit-to-data error, $E_{\text{threshold}}$, are set by the user. The fit-to-data error is evaluated employing the RSS, Eq. (8), and the complexity of each individual is quantified by counting the number of candidate terms in it. For the first-generation round, the fitness is measured just according to the RSS value. Thereafter, the RSS is normalized by the maximum RSS magnitude belonging to the set of the best individuals associated with the previous generation. Moreover, the number of active terms in each individual is counted and normalized according to the most complex individual in $G_b$ of the previous round. The fitness measure is defined as the normalized fit-to-data error either shrunk or expanded by the normalized complexity raised to power $\alpha$. The power has a value of zero until the generation round becomes equal or more than $N_{threshold}$ when $\alpha$ becomes one. The next change occurs when the RSS gains a value less than $E_{threshold}$. By this time, the algorithm gradually increases the magnitude of $\alpha$ to promote the parsimony among the individuals. If abrupt increase in the RSS is observed, the algorithm adaptively halves the increment in $\alpha$. This process continues until the fitness function converges.

$$\text{RSS} = \sum_i \left( \dot{Z}_i - (\Theta_P^n)_{ij} \xi_j \right)^2 \qquad (8)$$

The active functions in each individual are counted and respective number saved in a vector designated by $\Gamma$ that each row indicates the number dedicated to the respective solution in the population. The maximum of the vector $\Gamma$ for the best individuals, $G_b$, is stored as the variable presented with $\max(\Gamma_b)$ and the maximum RSS of the best individuals is $\max(\text{RSS}_b)$. The fitness measure can be formulated as follows

$$\Upsilon_i = \frac{\text{RSS}(X_i)}{\max(\text{RSS}_b)} \times \left[ \left( \frac{\Gamma(X_i)}{\max(\Gamma_b)} \right)^\alpha \left( \frac{\sum_{j=1}^{T_i} \mathfrak{I}(X_{i,j})}{\max(\mathfrak{I}_b)} \right)^\tau \right] \qquad (9)$$

where $\Upsilon_i$ is the fitness function computed for $i^{th}$ individual and $\alpha$ depicts the power to which the normalized complexity rises. In addition to the definition of the complexity that is the number of candidate terms in an individual, one might argue that the complexity of each candidate term can also be of importance. Therefore, a correction to Eq. (9) is suggested in which the number of operators appeared in each function term is counted, including mathematical operators and basis functions appearing in the function set such as sin, cos, etc. and subsequently added them together and divided by the number of functions being present in an individual $i$, which is designated by $\mathfrak{I}(X_{i,j})$ where $X_{i,j}$ is $j^{th}$ function term of the individual $i$. In Eq. (9), $T_i$ stands for the number of candidate terms in an individual $i$. The magnitude of this parameter is subsequently normalized by the maximum corresponding amount in the set of best individuals $\max(\mathfrak{I}_b)$. This correction is multiplied to the relationship presented in Eq. (9). The reason to count the number of operators instead of characters in each function term is that the algorithm differentiates "*$xt$" and "sin$t$" owing to different number of characters in them, i.e., 3 and 4 respectively. Therefore, the algorithm prefers the former expression, while both expressions are of the same complexity from a mathematical point of view. Thus, counting the number of operators in each candidate term outperforms and treats both above terms the same. The magnitude of $\tau$ in Eq. (9) is of a very small size $\tau \ll 1$ and the suggestion is to choose it in range $(0, 0.1]$.

In addition to the adaptive strategy just described, there are available statistical measures for model selection, derived based on Information Theory, which are trade-offs between the goodness of fit-to-data error and individual complexity. Two of these methodologies are also adapted to act as the driving force of the genetic programming developed in this study: (*i*) the Akaike information criterion (AIC) presented in Eq. (10) [42] and (*ii*) the Bayesian information criterion (BIC) in Eq. (11) [23]. The complexity of an individual is evaluated according to the number of candidate terms in it, designated by *d*. The fit-to-data error assessment is also carried out using the mean of RSS. Finally, $N_o$ in these formulations shows the number of observations.

$$\text{AIC} = N_o \log\left(\frac{RSS}{N_o}\right) + 2d \tag{10}$$

$$\text{BIC} = N_o \log\left(\frac{RSS}{N_o}\right) + d\log(N_o) \tag{11}$$

### 4.3. Hard-thresholding ridge regression

After constructing the best individual set $G_b$, the data belonging to this set are sent to an embedded subroutine in the AI system, which is called the hard-thresholding ridge regression [14]. In addition to sparsity promoted by the fitness measure and sparse regression, this subroutine also promotes sparsity of the members of $G_b$ using a hard-thresholding technique in which candidate functions with the following properties are chosen for removal provided that the effect of their removal on the corresponding RSS is neglectable.

$$\mathcal{K}_i^n = \left\{ F_k \in \{F_1, F_2, \ldots, F_{T_i}\}_n \mid \xi_k \leq \sigma \xi_{max} \right\} \tag{12}$$

where $T_i$ is the number of functions in the individual *i* of the generation round *n*, $F_k$ is a candidate function in the individual and $\xi_k$ is its coefficient obtained from the ridge regression while $\xi_{max}$ is the maximum value of coefficients related to that individual and $\sigma$ is a constant defined by the user that takes value $\sigma \ll 1$, e.g., 0.0001. The set of possible candidate functions for removal is determined for each individual under consideration before the effect of their removal on the RSS is estimated. If the change in the computed RSS after their removal is neglectable compared to the initial one, the candidate terms are removed from the individual *i*. It is noted that if the set of candidate terms for removal consists of more than one function and the removal of all of them in one move is not allowed, the algorithm automatically considers one by one removal scenario.

## 4.4. Sparse regression to determine unknown coefficients

Consider that the function **f** is unknown, Eq. (3), and no prior knowledge of the system is available. One might suggest considering a space of all possible functions while searching to discover the equations of motion based on TS data collected from a given multibody system [14]. Implementing this idea, the time-series datasets is sampled at a sequence of time points of the size *m* and two following matrices are constructed

$$\mathbf{Z} = \begin{bmatrix} \mathbf{z}^T(t_1) \\ \mathbf{z}^T(t_2) \\ \vdots \\ \mathbf{z}^T(t_m) \end{bmatrix}, \quad \dot{\mathbf{Z}} = \begin{bmatrix} \dot{\mathbf{z}}^T(t_1) \\ \dot{\mathbf{z}}^T(t_2) \\ \vdots \\ \dot{\mathbf{z}}^T(t_m) \end{bmatrix} \quad (13)$$

The state of the system at time $t_j$ is represented as vector of $\mathbf{z}(t_j) = [z_1(t_j) \quad z_2(t_j) \quad \ldots \quad z_n(t_j)] \in \mathbb{R}^n$ and *T* is the operation sign of matrix transpose. By this time, a number of possible functions are generated using the genetic programming (GP) algorithm. Given the set of function terms, a library matrix, $\mathbf{\Theta}_P^n$, is constructed that contains the values of the candidate terms at discrete time steps, as is demonstrated below

$$\dot{\mathbf{Z}} = \mathbf{Z}_t = \mathbf{\Theta}_P^n(\mathbf{z},t)\boldsymbol{\xi} \quad (14)$$

$$\mathbf{\Theta}_P^n(\mathbf{z},t) = \begin{bmatrix} | & | & & | & \\ G_P^1 & G_P^2 & \cdots & G_P^k & \cdots \\ | & | & & | & \end{bmatrix}^n \quad (15)$$

$G_P^k$ is associated with $k^{th}$ member of the individual *P* in the $n^{th}$ population. The $k^{th}$ column of the library matrix contains the values of the candidate term *k* of the individual *P* in the population. To clarify the procedure, consider the $j^{th}$ member of an individual is of the form $\frac{z_2(t)\sin z_3(t)}{t}$, the following vector places in the $j^{th}$ column of the library matrix

$$\left(\mathbf{\Theta}_P^n(\mathbf{z},t)\right)_j^T = \begin{bmatrix} \frac{z_2(t_1)\sin z_3(t_1)}{t_1} & \frac{z_2(t_2)\sin z_3(t_2)}{t_2} & \cdots & \frac{z_2(t_m)\sin z_3(t_m)}{t_m} \end{bmatrix} \quad (16)$$

There is a large number of candidates generated by the GP algorithm to build the entries in this library matrix. The coefficients, $\boldsymbol{\xi}$, determine which function terms are active and to what extent they, together, are close to the true form of the function **f**. Here, the number of equations is considered more than unknown coefficients that is equal to the number of candidate functions existing in each individual. The set of equations are fit to time-series data using a regression method like the least squares method that provides estimates of the unknowns by minimization of the objective function $\left\|\mathbf{\Theta}_P^n\boldsymbol{\xi} - \dot{\mathbf{Z}}\right\|_2^2$ for a set of algebraic equations $\mathbf{\Theta}_P^n\boldsymbol{\xi} = \dot{\mathbf{Z}}$. Commonly, resulting estimated coefficients are nonzero, which makes the interpretation of the discovered model challenging when the size of the system is large, in particular. Moreover, this solution does not fulfil the constraint on the form of the selected function to be parsimonious. Therefore, there is a need for regularizing the minimizing process to promote sparsity in the set of unknown coefficients. This means that a sparse vector of $\boldsymbol{\xi}$ should have been sought to end up with a parsimonious model for a given multibody system, which is why one may be interested in regularizing regression methods such as sequential threshold ridge regression, the least absolute shrinkage and selection operator (Lasso), and elastic-net approach.

A generalized Tikhonov functional for a problem of the form $\mathbf{\Theta}_P^n\boldsymbol{\xi} = \dot{\mathbf{Z}}$, for which there is not a well-defined inverse operator $(\mathbf{\Theta}_P^n)^{-1}$ such that $\boldsymbol{\xi} = (\mathbf{\Theta}_P^n)^{-1}\dot{\mathbf{Z}}$, can take the following relationship [34, 35]

$$T_\lambda(\boldsymbol{\xi};\dot{\mathbf{Z}}) = \mathcal{L}(\mathbf{\Theta}_P^n\boldsymbol{\xi},\dot{\mathbf{Z}}) + \lambda J(\boldsymbol{\xi}) \tag{17}$$

$\lambda$ is the regularization parameter, $J$ is called the penalty (regularization) functional while $\mathcal{L}$ represents fit-to-data functional, which quantifies how close the prediction $\mathbf{\Theta}_P^n\boldsymbol{\xi}$ is to the observed data $\dot{\mathbf{Z}}$. The $\ell_2$ space norm is the most familiar expression for the latter one, which is expressed as follows

$$\mathcal{L}(\boldsymbol{\xi},\dot{\mathbf{Z}}) = \left\|\mathbf{\Theta}_P^n\boldsymbol{\xi} - \dot{\mathbf{Z}}\right\|_2^2 \tag{18}$$

The regularization functional can take different forms according to the methodology that is chosen to incorporate a priori information. The ridge regression method uses a penalty functional of the form $\ell_2$ norm to penalize the unknown coefficients, that is $J(\boldsymbol{\xi}) = \|\boldsymbol{\xi}\|_2^2$. Differentiating the functional $T_\lambda$ with respect to unknown coefficients, $\hat{\boldsymbol{\xi}} = \text{argmin}_{\boldsymbol{\xi}} T_\lambda$, one can obtain a closed-form relationship for the vector of coefficients minimizing the functional $T_\lambda$. The least absolute shrinkage and selection operator [43] that is also known as lasso that employs an $\ell_1$ constraint on the absolute magnitudes of the coefficients, expressed by $J(\boldsymbol{\xi}) = \|\boldsymbol{\xi}\|_1$. This constraint leads to the shrinkage of the coefficients, promoting sparsity in coefficients. The optimization problem whilst using the lasso is convex but has a singularity at zero. The gradient decent method is employed to solve the optimization problem associated with the use of the lasso method along with a soft thresholding.

In the case of highly correlated variables, using the lasso leads to inconsistent outcomes and the coefficients estimated for different Lagrange multipliers demonstrate erratic paths. A quadratic penalty is believed that can help the lasso cope with this situation as it happens in the elastic-net regression, which is defined by $J(\boldsymbol{\xi}) = \alpha\|\boldsymbol{\xi}\|_2^2 + (1-\alpha)\|\boldsymbol{\xi}\|_1, \alpha = \frac{\lambda_2}{\lambda_1+\lambda_2}$ [44]. It is worth mentioning that the elastic net has a singularity when $\alpha = 0$, while is convex generally; otherwise, it is strictly convex $\alpha \in (0, 1]$. This regression method shows the characteristics of both the lasso and ridge regression methodologies. Given a dataset $(\boldsymbol{\Theta}_P^n, \dot{\mathbf{Z}})$, a new dataset is defined as $(\boldsymbol{\Theta}_P^{n*}, \dot{\mathbf{Z}}^*)$ to convert the elastic net regression to a form of the lasso that one already knows how to solve.

## 5. Hybrid dynamical multibody systems

Among natural and engineering systems, there are hybrid dynamical systems that operate in several dynamics modes switching from one to another over time [24, 26]. The transition from one mode to another can take place smoothly or with abrupt changes due to events such as impact, switching, and frictional sliding. Gaining knowledge on operating modes, transition and reset maps of such dynamical systems plays a key role in system identification and parameter estimations [45]. Moreover, discovering the nonlinear dynamics of such systems is more complicated than that of the usual smooth dynamical systems operating in just one dynamic mode due to the existence of several submodels, switching edges (boundaries), and switching sequence [24].

Consider a dataset of input and output data, $\mathcal{D} = [\mathbf{Z} \ \dot{\mathbf{Z}}]$, collected from a hybrid dynamical system. One needs to know how many modes this system operates in and what respective submodels (substructures) are. The sequence by which the system switches between modes is also of importance for system identification. The events that cause the dynamic system to switch from one mode to another can be a function of both time and system state. One can construct a new dataset of clusters $\overline{\mathcal{D}} = \left\{\aleph_T^k(\mathbf{Z}_k, \dot{\mathbf{Z}}_k)\right\}_{k=1}^m$ where $\aleph_T^k$ represents the $k^{\text{th}}$ cluster of the training data and $m$ is the number of time points. Each cluster, $\aleph_T^i$, consists of $K$ data belonging to $\mathcal{D}$ that are nearest in terms of the Euclidean distance to the cluster's centroid, $C_T^i$, determined at each time point $t_k$. These clusters are obtained using the $K$-nearest neighbors algorithm. Having a look at the unknowns of this problem, one can say this problem is not well-posed. Several systematic ways are proposed to define the possibly well-posed problem of system identification of a hybrid dynamical system, one of which is to consider a fixed number of dynamic modes and try to search for respective sub-models. One can, thus, redefine the problem as follows. While number of submodels, $s$, is known, discover not only models associated with each mode $\left\{(\mathbf{f}_j)\right\}_{j=1}^s$, but also the sequence of switching between the submodels $\mathbf{S} = \{S_k\}_{k=1}^M \in [s]^m$ by minimizing the following functional [23]

$$\min_{\{\mathbf{f}_i \in \mathcal{F}\}_{j=1}^s, S \in [s]^M} \frac{1}{m} \sum_{k=1}^m \sum_{i=1}^K \ell\left(\dot{\mathbf{z}}_{i,k} - \mathbf{f}_{S_k}(\mathbf{z}_{i,k})\right) \mid [\mathbf{z}_{i,k} \ \dot{\mathbf{z}}_{i,k}] \in \aleph_T^k \qquad (19)$$

where $\ell$ is a fit-to-data functional. This optimization problem is nontrivial because of containing not only continuous but also integer variables (function indexes). It might be interesting to know that there are two extreme solutions for such a problem. The trivial solution consists of as many substructures as data points (overfitting). One can also choose $s = 1$ and easily estimate a single submodel, which leads to high error. The above definition of the problem and this latter extreme solution motivate our approach that is described as follows [23].

- It is assumed that the number of substructures is one, $s = 1$. The first submodel is discovered with the dataset of the cluster $i$, $\overline{\mathcal{D}}^i = \{\aleph_T^i\}$, using the evolutionary symbolic sparse regression module that satisfies the functional minimization in Eq. (19) for the cluster dataset at hand. The discovered model is, in turn, added to the set of submodels $\{(\mathbf{f}_j)\}_{j=1}^S$. This solution is not an extreme solution as the respective error is minimized and balanced with the complexity of the distilled model, provided that this respective cluster does belong data associated with just one mode of the system dynamics. The respective element of the switching sequence vector, $\mathbf{S} = \{S_k\}_{k=1}^M$, gains the value 1, i.e., $S_i = 1$. So, one looks for a cluster $\aleph_T^i$ that does not convey any events in its dataset.
- The next cluster is assessed to see whether it belongs to the previously discovered submodel or not. If yes, $S_{i+1} = 1$; otherwise, the algorithm is run to discover the corresponding motion equations for time-series data $\overline{\mathcal{D}}^{i+} = \{\aleph_T^{i+1}\}$ and the number of submodels increases by one, i.e., $s_{new} = s_{old} + 1$, and $S_{i+1} = 2$. This process continues until every cluster is assigned to a substructure and the $m$ members of the switching sequence vector is filled.

The procedure is elaborated in more detail in the following. Assume $\aleph_T^j$ is the $j^{th}$ cluster and does not convey any events in its data set. This cluster consists of $K$ data points nearest to $C_T^j$ that is the centroid of the cluster $j$. Respective submodel is sought and the model discovered is indicated by $\mathbf{f}_j$ that is an individual consisting of several function terms, i.e., model features. The algorithm stores this substructure and its features in the set of (i) submodels $\{(\mathbf{f}_j)\}_{j=1}^S$ and (ii) features (approved basis functions) $\{(\beth_j)\}_{j=1}^\vartheta$. Features include both candidate terms and constant values of a selected individual. The next cluster, $\aleph_T^{j+1}$, is treated as follows. From a physical point of view, it is assumed that the mathematical structure of submodels does not vary to large extent. In other words, the equations of motion are the same somewhat. However, new function terms can be added or removed and coefficients appearing in the equations can alter. For example, the equations of motion (EoM) of a system subjected to friction does not change with the occurrence of switching in friction behavior. Another example can be a system to which an external body contacts during its motion. This mechanical contact can be regarded as an external force that is added to the main EoM of the multibody system. This viewpoint is embedded in the suggested procedure.

a) Consider a new cluster. The possibility that this new cluster is governed by any of the previous discovered submodels is assessed. If yes, the respective submodel is known and respective element of the switching sequence vector is determined.

b) If the step (*a*) does not work, there are two possibilities one of which requires searching for a new submodel and the other suggests the existence of at least one event in the corresponding cluster. The first possibility is assessed by searching to obtain a new substructure. Features $\{(ℶ_j)\}_{j=1}^{\vartheta}$, in which $\vartheta$ is the number of distinct features, are used as a prior knowledge and are integrated in the GP algorithm. In this study, such features are randomly taken by the algorithm whenever candidate terms produced by the crossover operation are of large size (depth), which is computed by counting terminals and functions in the tree, and incompetent. The other available option to take those features into account is to add them to the function set.
c) The possibility of existence of events in the data associated with the cluster in *b* is also investigated. A technique to identify the data index associated with the switching boundary is applied to such a cluster. The vector of data of this cluster is split into two regions of sizes with indexes [1: Π-1] and [Π: *K*] [26]. An optimization problem is defined in which the sum of residual error from its local mean is minimized based on the variable Π, which results in a time index close to the event, designated by $t_s$. This procedure is applied to the simulated data, obtained by using the discovered model, and the validation dataset. When the switching time, $t_s$, is estimated, discrete time steps in its vicinity are also tested for possible update to improve the accuracy according to new information gained by determining $t_s$, one divides the cluster into two according to $t_s$ and search to discover either model.
d) The validation of the submodels is done using the standard techniques relying on an additional and independent data set, called a validation set. At the end, the submodel of a cluster with smallest validation error is retained. If it is not possible to obtain a validation set due to, for example, the waste of data and finite sample size, one can employ the *K*-fold cross-validation approach by dividing the training data into *K* parts and, subsequently, using each part once as the validation set and the others for training. Finally, the RSS is averaged for all validation sets (*K* sets) utilized in the study [23, 43].

### 5.1 Clustering

Dividing the time-series data into several clusters to solve the problem of identifying the hybrid dynamical system, Eq. (19), the clustering is described in this section. The overall goal of the clustering used in this article is to divide a set of data points into several groups or classes based on some measure of similarity between objects such as the Euclidean distance to the cluster centroid. *K*-nearest neighbors algorithm is employed to find a group of *K* similar measurements in the training set. Data-driven coordinates are considered according to the state space of a given multibody system, which is of the form $[\mathbf{Z}, \dot{\mathbf{Z}}]$ and of the size depending upon the number of system states. If one considers a system of two states, the states of the system can be represented in a 2D plot. The *K*-nearest neighbors algorithm is applied to the set of data and the clusters are found for each time point, $t_j$, and subsequently the centroid of each cluster, $C_T^j$, is obtained. The corresponding validation cluster, $C_V^j$, is also determined finding *K* data points in the validation set closest to the centroid of training cluster $C_T^j$. The validation data of this validation cluster are later used to validate the substructure discovered for the respective cluster [23, 26].

Defining data-driven coordinates can be challenging. One can argue that the best data-driven coordinates are those of the system states. The other suggestion is to consider all coordinates, whose data are available, and build data-driven redundant coordinates. Choosing either displacement coordinates or velocities as data-driven coordinates constitute other possibilities. Now, the question raised is that which data-driven coordinates can be more efficient. Just to give an example, consider the classical bouncing ball problem in which a ball is released from a height. It collides with the ground and rebounds. During the time the ball is in contact, the equation of motion is governed by a different submodel compared to the time during which it is in the free motion mode. Two events are recognized that occur at times the ball impacts the ground, i.e., contact commencement, and leaves the ground during the rebound, i.e., the end of contact phase. If one wants to differentiate these two motion modes, considering the displacement coordinate is enough and is more efficient than that of system states including both velocity and displacement variables [46]. As the second example, one can consider the sliding motion of a body against a fixed surface subjected to friction. It is assumed that friction coefficient varies with the velocity based on the following formula [47].

$$\mu(v_{rel}) = \begin{cases} \left(c_d + (c_f - c_d)\exp(-\xi(|v_{rel}| - v_0))\right)\text{sgn}(v_{rel}) & |v_{rel}| > v_0 \\ \left(c_f - \dfrac{c_f}{v_0^2}(|v_{rel}| - v_0)^2\right)\text{sgn}(v_{rel}) & |v_{rel}| \leq v_0 \end{cases}, \quad (20)$$

As is observable from Eq. (20), friction model switches from one model to another once the magnitude of the relative velocity reaches $v_0$. Therefore, it can be inferred that a better data-driven coordinate defined for such a problem consists of just velocity coordinate. Another problem is the multibody dynamics of the knee joint, in which the ligament force imposed on the knee joint has the following form [48]

$$f(\epsilon) = \begin{cases} \dfrac{k\epsilon^2}{4\epsilon_1} & 0 \leq \epsilon \leq 2\epsilon_1 \\ k(\epsilon - \epsilon_1) & \epsilon > 2\epsilon_1 \\ 0 & \epsilon < 0 \end{cases}, \quad (21)$$

The ligament force switches between three models based on the value of the ligament strain, $\epsilon$, that varies with the relative displacements of the knee components. One can again claim that an efficient way to construct clusters is based on the displacement coordinates of such a multibody system. Therefore, it is suggested to consider different types of data-driven coordinates, enabling several feature extractions from the time-series data of a given multibody system.

# 6. Demonstrative examples and predefined parameters of the algorithm

Three illustrative case studies are considered in this study to demonstrate the capability of the developed method: (*i*) a two DoF spring-mass system; (*ii*) crank-slider mechanism; and (*iii*) sliding mass subjected to friction. The programming codes to implement the method are all written in MATLAB (R2017a) and the algorithm is run on a 1.80 GHz personal computer with Intel(R) Core(TM) i5-8250U CPU. The maximum size (depth) of each tree is considered 15. $N_{threshold}$ is initially set 50 and $E_{threshold}$ is set depending on the order of observed variables. After the initial test, those magnitudes can be adjusted by the user. The code does not stop searching before number of generations is less than 50 and the scaled mean RSS (scaled by the order of observations) greater than 1e-3. $\vartheta$ in Eq. (7) is also set 5. The number of data points, *K*, considered in *K*-nearest neighbors algorithm is determined with respect to the size of data. The constant values are randomly selected to be in range of [-100 100], adjustable. The maximum number of function terms considered in each individual is 20 and the size (depth) of each function tree is 5 in the first round of generating the population and later can increase to 15. The percentage of population size that is dedicated to the reproduction operation is $\varepsilon\% = 20\%$, Fig. 4. The current version of the algorithm can handle multibody systems with 50 system coordinates, which can be adjusted to any number upon need.

# 7. Results and discussion

This section aims at reporting results obtained for each of three case studies investigated in this article. In the following, three subsections are allocated to the analysis of the demonstrative examples, in which the capability of the developed method to discover physics-based models, its robustness while the data is corrupted with noise, and the effect of data length on the distilled governing equations are presented. Efficiency of several time-differentiate approaches are investigated and reported in Section 7.1. Hybrid dynamical systems and the method presented in Section 5 are detailed in Section 7.3. Three fitness measures are assessed in this section and their general performances are discussed.

### 7.1. Demonstrated example *i*: A two-DoF spring-mass system

Forced vibration of a two degree-of-freedom (DoF) mass spring system is considered in this section, as is demonstrated in Fig. 7. External harmonic forces, i.e., $f_1(t) = -200\sin 2t$ and $f_2(t) = 100\sin(5t + \pi/3)$, are respectively applied to the bodies with masses of $m_1$ = 10 kg and $m_2$ = 5 kg. Moreover, stiffnesses of springs are $k_1$ = 200 NM$^{-1}$, $k_2$ = 300 NM$^{-1}$, and $k_3$ = 200 NM$^{-1}$. State variables of the system are depicted by $\mathbf{z}^T = [z_1 \quad z_2 \quad z_3 \quad z_4] = [x_1 \quad x_2 \quad \dot{x}_1 \quad \dot{x}_2]$. Time-series data are collected from this multibody system while considering several initial conditions such as $\{0, 0, 0, 0\}, \{1, -0.2, 0, 0\}$. Function and terminal sets are $\{+, \times, \div, \sin, \text{sqr}, \text{abs}, \exp, \text{sgn}\}$ and $\{t, z_1, z_2, z_3, z_4, c, 1\}$. The population size is 600 and the number of generations is 100 for all experiments. Two datasets are considered for each experiment, namely training set and validation set. Results reported for each experiment are an average of several runs of the algorithm (at least 5 times).

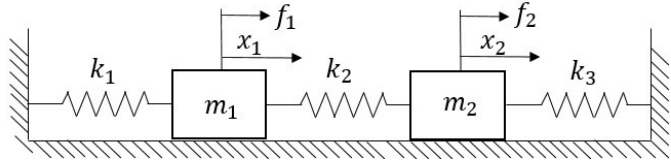

Fig. **7**. A two DoF dynamical system.

Three fitness functions defined in Section 4.2.2 are employed to evaluate their performances. According to experiences carried out in this study, it can be concluded that the fitness relationship defined in Eq. (9) outperform the others in terms of balancing the parsimony and accuracy. However, the Akaike and Bayesian information criteria do not need parameters to adjust compared to the one introduced in Eq. (9) that requires predefining two parameters by the user. The resulting model that satisfies a balance between the fit-to-data accuracy and complexity is discovered using the suggested algorithm and presented in Eq. (22). Using the fitness measure of Eq. (9), $N_{threshold}$ is considered 50 while the number of run 100 and $\alpha$ increases upto 2. The accuracy of the model, determined based on the training set, is RSS = 0.005 and the number of function terms in the individual 6. The error based on the validation set is also estimated, which is RSS = 0.112. The other experiment is to consider that one has a prior knowledge of the force vector applied to the mass 2. This input data, i.e., $u_1$, is added to the terminal set as $\{t, z_1, z_2, z_3, z_4, c, 1, u_1\}$. The resulting model is presented in Eq. (23).

$$\dot{z} = \begin{pmatrix} 1.0000 z_3 \\ 1.0000 z_4 \\ 20.0000 \sin(t+t) + 30.0000 z_2 - 50.0001 z_1 \\ -19.9903 \sin(-4.9981 t - 1.0479) + 59.9493 z_1 - 100.0189 z_2 \end{pmatrix} \quad (22)$$

$$\dot{z} = \begin{pmatrix} 1.0000 z_3 \\ 1.0000 z_4 \\ 20.0000 \sin(t+t) + 30.0000 z_2 - 49.9997 z_1 \\ 0.2001 u_1 + 59.9518 z_1 - 100.0180 z_2 \end{pmatrix} \quad (23)$$

Furthermore, the robustness of the developed method is evaluated by adding white Gaussian noise with zero mean to the time-series data and the algorithm is employed to identify the system. Several signal-to-noise ratios (SNR) are considered, that are, 15, 20, 25, 30, and 40 dB. Signal-to-noise ratio can be defined as $\text{SNR} = \frac{P_{signal}}{P_{noise}}$ where $P_{noise}$ and $P_{signal}$ are the power of the background noise and signal, respectively. As can be seen in Table 1, increasing noise in the input signals cause the error to increase. Moreover, the structure of model undergoes changes itself either in coefficients or in function terms. For example, when SNR is 20 dB, one terms is removed from the resulting equations of motion, i.e., $\sin(ct + c)$. The same noisy data, used for the models listed on the left side of the table, are denoised to some extent using the total variation technique and respective governing equations are obtained. The respective errors show a decrease compared to those obtained from intact data. Therefore, the suggestion is to first denoise time-series data to smooth just very sharp changes observed in signal and, in turn, apply the algorithm.

Table 1. Effect of noisy signals on error and complexity of the discovered models.

| SNR (dB) | Intact data | | Smoothed data | |
|---|---|---|---|---|
| | $\sqrt{\dfrac{RSS}{N_0}}$ | Complexity (term no.) | $\sqrt{\dfrac{RSS}{N_0}}$ | Complexity (term no.) |
| 40 dB | 1.33 | 6 | 0.75 | 6 |
| 30 dB | 3.97 | 6 | 2.21 | 6 |
| 25 dB | 6.16 | 6 | 3.13 | 6 |
| 20 dB | 10.17 | 5 | 5.03 | 6 |
| 15 dB | 8.29 | 12 | 7.58 | 6 |

The variation trend of the mean error, $RSS/N_o$, with respect to the complexity, $\Gamma(X_b)$, of the best individual in each generation is also considered and plotted in Fig. 8. The fitness function used in this section is the one presented in Eq. (9) while $\alpha$ increases up to 2. The number of function terms in the individual decreases from 12 to 6 in a nonlinear fashion. When the complexity drops from 11 to 6, the error rises two-fold, that is, ~0.1. The best solution occurs when the complexity turns out to be six and the mean error 5e-6. It is worth mentioning that right after diminishing the complexity from 7 to 6, the error is too much higher than 5e-6. However, what algorithm does is to keep the complexity the same and try to modify the constants, present in the function arguments, in order to improve the error.

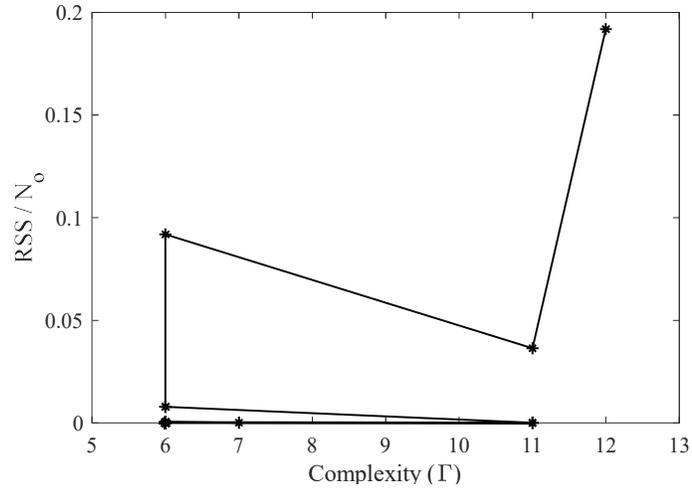

Fig. **8**. Mean RSS with respect to the model complexity for the last 65 generation rounds of the algorithm. The mean RSS and complexity for the first generation are 62.785 and 28, respectively.

The influence of the length of time-series data is also investigated on the algorithm performance. Several sampling sizes of input data is considered, i.e, 20, 100, 167, 250, 500 and 1000. The corresponding models are obtained, and the algorithm does discover the model efficiently. The error and complexity of the models discovered for each data sampling is presented in Table 2. A question might be raised that why the error associated with the training set with 20 sampling data is greater than that associated with the validation set. The reason is the first solution of this problem leads to an error of 2e-5, which seems very promising, but the error estimated from the validation set shows a terible result, that is, 2.8e4 with a dissapointing complexity of 17. The occurence of overfitting can be infered . Our strategy is to use both training and validation sets to do system identification such that the fitness function is computed based on the validation set and the sparse regression with the training set while the number of candidate terms in each individial is counted to determine the complexity of each individual. Using this methodology, the algorithm does not undergo overfitting and outcomes are reported in Table 2 for the experiment with 20 sample data.

Table 2. A comparison of the error and complexity of the discovered models with multiple sampling sizes.

| Sample size | $e_T = \dfrac{RSS_T}{N_T}$ | $e_V = \dfrac{RSS_V}{N_V}$ | Term no. |
| --- | --- | --- | --- |
| 1000 | 5.0e-6 | 1.1e-4 | 6 |
| 500 | 1.6e-6 | 9.4e-6 | 6 |
| 250 | 6.8e-6 | 4.1e-4 | 6 |
| 167 | 3.9e-4 | 0.0250 | 6 |
| 100 | 1.5e-4 | 3.8e-4 | 6 |
| 20 | 0.0370 | 0.0069 | 6 |

This demonstrative example is also used to study the efficiency of three approaches for time derivations of noisy data. A Gaussian white noise with zero mean is added to the data and time-series data available is limited to the displacement vector Fig. 9a. Therefore, one needs to obtain velocity and acceleration vectors from the time derivative of displacement. As is plotted in Fig. 9b, the finite difference method (FDM) does not provide us with efficient velocity data from the displacement. The Savitzky-Golay filter does a better job compared to FDM; however, the resulting velocity data accommodates a large error itself, due to the noise existing in the displacement vector. The other methodology employed to do time differentiation is the total variation regularization that Fig. 9c demonstrates a very good agreement between the numerically computed data (without noise) and those obtained from the TV method. The discovered model for the full set of data without noise and the incomplete noisy data are also given in Eqs. (24) and (25) for the sake of comparison.

$$\dot{\mathbf{z}} = \begin{pmatrix} 1.0000 z_3 \\ 1.0000 z_4 \\ -19.8900 \sin(-1.9998 t) + 27.9464 z_2 - 48.5877 z_1 \\ 20.0511 \sin(-4.9930 t - 4.2357) + 60.8927 z_1 - 100.7832 z_2 \end{pmatrix} \quad (24)$$

$$\dot{\mathbf{z}} = \begin{pmatrix} 1.0000 z_3 \\ 1.0000 z_4 \\ -20.0002 \sin(-2.0003 t) + 30.0308 z_2 - 49.9949 z_1 \\ -19.9945 \sin(-5.0074 t - 1.0361) + 59.3456 z_1 - 99.9539 z_2 \end{pmatrix} \quad (25)$$

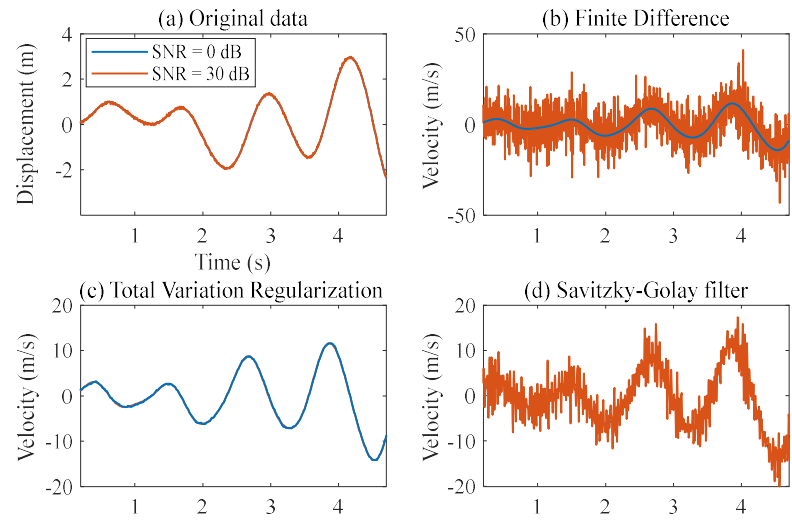

Fig. **9**. Comparison of three methods to differentiate noisy data.

One might finally discuss the extrapolative capability of the developed algorithm. The validation set constructed in the present study consists of data outside of the span of training dataset both in terms of initial conditions under which the validation data are produced as well as time interval of the physical event. According to results reported in this section, it can be concluded that one of the advantages of the presented methodology compared to machine learning methods is its extrapolative capability as the learning machines are fundamentally interpolative. The algorithm can produce physics-based models for given multibody systems, which work well outside of the probability distribution with which the system is trained. The interpretability of discovered models can also be highlighted, and one can understand the underlying mechanisms in systems of interest. In addition, the algorithm can be trained and generalized, and one can, in turn, use it for input data associated with any initial conditions [14, 17, 49].

### 7.2. Demonstrated example *ii*: Crank-slider mechanism

A crank-slider mechanism consists of four links with three revolute joints and one sliding, as is demonstrated in Fig. 10. The linear movement of the slider is driven by the external torque applied to the crank, which leads to an angular velocity of the crank. The link AB of the linkage considered in this study is of 1 m length while the length of the link BC is 1.5 m. The angular velocity of the crank is assigned to be 20 rad/s and the linkage begins to operate when the link AB constructs the angle zero with *x*-axis. The reason of choosing such a mechanism is the equation of motion has a rational form that is not trivial to be discovered owing to its complex form. Moreover, there are several constants appearing as the arguments of motion equation, which require to get them all estimated during the searching process, and it is not possible to identify them using the sparse identification of nonlinear dynamic system unless one knows the model and just tries different values for the constants one by one to minimize the fit-to-data error. Function and terminal sets chosen in this experiment are $\{+, \times, \div, \sin, \cos, \text{sqr}, \text{abs}, \text{sgn}\}$ and $\{t, z_1, c, 1\}$. The population size is 800 and the algorithm continues to regenerate populations until the fitness measure, Y, converges while the power $\alpha$ varies from 1 to 2. The governing motion equation of slider-crank linkage is discovered as are given in Eqs. (26) and (27). The latter is doing as good as the former formula but with different argument constants. Two datasets are considered for each experiment, namely training and validation sets. Results reported for each experiment are an average of several runs of the algorithm (at least 3 times).

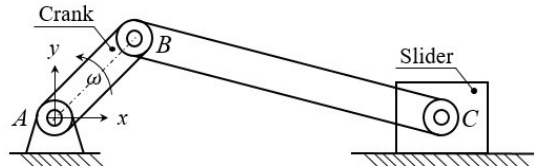

Fig. 10. Crank-slider mechanism (initial value: $z|_{t=0} = L_1 + L_2$).

$$\begin{aligned}\dot{z}_1 &= -20.0018 \sin(20.0010t) \\ &- 20.0047 \frac{\sin 20.0210t \sin(19.9054t + 1.5639)}{\sqrt{2.3720 + \sin(-20.0149t)\sin(20.0311t)}}, \quad (\dot{z}_1 = \dot{x}_C)\end{aligned} \quad (26)$$

The other format of the governing equation discovered by the developed method is given below, whose second term has a different structure in appearance with dissimilar constant values compared to one in Eq. (26). However, a formulation like the respective analytical equation can be obtained after some mathematical manipulation using the trigonometric addition and subtraction formulae. The mean RSS is 4.95e-3 and the number of terms is 2. The number of population run is 233. It is worth noting that Eq. (27) is less complex than Eq. (26) due to the number of present constants and operators in its second expression.

$$\begin{aligned}\dot{z}_1 &= 19.9993 \sin(-19.9983t) \\ &+ 14.2618 \frac{\sin(-39.9993t)}{\sqrt{3.5586 + \cos(-39.9969t)}}, \quad (\dot{z}_1 = \dot{x}_C)\end{aligned} \quad (27)$$

The advantage of the developed method is to successfully generate various forms of the rational functions likely contributing to the governing equation of the multibody system without any need to change the methodology as is done in Mangen et al. [19]. A limitation observed when using the suggested methodology is that when the quotient squared of the connecting rod to the rotating crank, $\|\overrightarrow{BC}\|/\|\overrightarrow{AB}\|$, is much greater than one being the maximum value that $(\sin ct)^2$ or $\cos ct$ can gain, the error of selecting a false magnitude for the sine or cos function argument in the denominator is very low when being compared to the other term in summation and it is likely that the algorithm does not show a required sensitivity to modify its amount according to our experiments.

### 7.3. Demonstrative example *iii*: Hybrid dynamical system: sliding mass subjected to friction

A hybrid dynamical system is also considered in this study, which is the forced vibration of a mass-spring system subjected to Stribeck friction, Fig. 11. Stribeck friction model accounts for the negative damping characteristic of friction. This model demonstrates a discontinuity problem at zero velocity and does not account for stick-slip friction. Bengisu and Akay suggested a friction formula, Eq. (20), to account for Stribeck effect, to resolve the discontinuity, and to capture stick-slip phenomenon [47]. The reasons to choose this demonstrative case study are that this is a hybrid system and there is a variety of function terms in the model such as signum, exponential, polynomial, and trigonometric functions. For the sake of state-space representation, the state variables are shown by $\mathbf{z} = \begin{pmatrix} z_1 \\ z_2 \end{pmatrix} = \begin{pmatrix} x \\ \dot{x} \end{pmatrix}$ in which $x$ depicts the displacement vector of the system. Function and terminal sets chosen are $\{+, \times, \div, \sin, \text{sqr}, \text{abs}, \exp, \text{sgn}\}$ and $\{t, z_1, z_2, c, 1\}$, respectively. The body is of a mass 10 kg while the spring stiffness is 2000 Nm$^{-1}$ and the external force has a harmonic form $200\sin(2t)$ N. The friction parameters seen in Eq. (20) are $c_f = 0.15$, $c_d = 0.065$, g = 9.81 ms$^{-2}$, $v_0 = 0.1$ m/s, $\xi = -3$. Several initial conditions are considered and respective data are concatanated together to form time-series data used to train and validate the multibody system. The initial conditions for the training set are $\mathbf{z}|_{t=0} = (0, 0), (0.001, 0), (-0.001, 0)$ and those for the validation set $\mathbf{z}|_{t=0} = (0, 0.15), (-0.01, 0.02)$. The time interval of interest is in a range of [0, 5] s while the validation set covers a longer time period up to 10 seconds. The number of clusters is the same as time points at which data are collected $t = [t_1, \cdots, t_m]$.

According to the experiment setup, there are 250 time points and, subsequently, the same number of clusters are generated using the *K*-nearest neighbors algorithm. As the data is constructed by concatenating data associated with multiple initial conditions together, all data that occur at time points very close to the one of interest are averaged to determine the cluster centroid. For each cluster, a submodel is determined, which can be similar with some of other discovered models. The size of the first population is 800 and all previously discovered submodels are embedded in the population, but the other generations are of size 500. On the other hand, an array that is called the function memory is built, which includes function terms of successful, previous submodels, i.e., features. Discovering a new submodel for the next cluster, the function terms are added to the function memory provided that new function forms are found. These candidate terms are imported in the population once the size of newly generated expressions after crossover action becomes longer than 15. The members of the function memory are randomly chosen and replaced by the expression generated. One might note that not all of expressions of size more than 15 are passed to this process and just a number of them are randomly selected to go for either the substitution or redoing the crossover process. In addition, constant values of the function arguments, associated with those previously successful submodels, are stored in the other array that is named the constant-variable memory. The members of this memory are used for the crossover of the constant arguments.

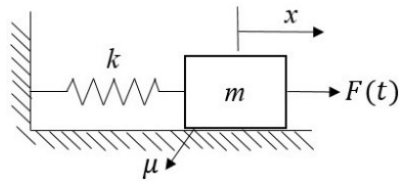

Fig. 10. Sliding body subjected to friction.

The method described in Section 5 is implemented to determine submodels and switching sequences. It is observed that a postprocessing is required as models found for the future clusters in time can outperform the previous ones. Therefore, all discovered models, some of which are the same especially after the final forms of submodels are obtained through the training, are tested for the first cluster to the last one. Two dynamic modes are recognized and respective submodels are named I and II, as is demonstrated in Fig. 12. It is worth noting that we found several models that fit to the data very well for each mode, but the algorithm promotes simplicity in their findings, i.e., the number of terms and complexity index. For example, the models distilled for clusters 119 and 121, which are given in Eqs. (28) and (29) and do fit the data for the first mode of the hybrid dynamical system, are the same in practice but the complexity of the former, i.e., 1.25, is less than the latter, i.e., 1.5. The model associated with the cluster 119 is, thus, preferred and is chosen as the submodel I for the whole system. The next function expression (submodel II) is written in Eq. (30). This latter governing equation can have been simplified more to reduce its complexity from 1.8 to 1.6 by removing the coefficient of $z_1$. In addition, some clusters behave oddly such as the cluster 35, as is seen in Table 3. The corresponding error that is the summation of errors obtained from the validation and training sets is high and not acceptable in comparison with others. Moreover, do none of the discovered models can be of any help to modify the performance. It implies that the cluster includes data belonging to at least two modes. The optimization technique discussed in Section 5 is employed to identify the data index associated with the switching point. The data is sorted with respect to time from the start to end and the analysis is carried out, but the technique does not work. In the next try, the data is sorted according to either displacement or velocity. The one obtained from the velocity obtains the switching point in the cluster as can be seen in Table 3. The mode is switched around the velocity 0.104 m/s. The finding process of switching point shows that the velocity parameter is determinative, and the clustering can be done based on merely velocity values rather than both system states, that are, displacement and velocity. Finally, one of the successful models discovered for this experiment is given in Eq. (31).

$$\{\sin 2t, z_2, z_2|z_2|, z_1\} \tag{28}$$

$$\{\text{sgn}(z_2)z_2^2, \sin 2t, z_2, z_1\} \tag{29}$$

$$\{\text{sgn}(z_2)\exp c|z_2|, \sin 2t, z_2, cz_1, 1\}, c = -2.9857, -78.245 \tag{30}$$

$$\dot{z} = \begin{cases} \begin{pmatrix} 1.000 z_2 \\ 19.998 \sin(t+t) - 199.983 z_1 - 0.639 \text{sgn} z_2 - 1.124 \text{sgn} z_2 \exp(-3.001|z_2|) \end{pmatrix} & |z_2| > v_0 \\ \begin{pmatrix} 1.000 z_2 \\ 19.993 \sin(2.000 t) - 199.925 z_1 + 146.734 z_2^2 \text{sgn}(z_2) - 29.391 z_2 \end{pmatrix} & |z_2| \le v_0 \end{cases} \tag{31}$$

Table 3. Examples demonstrating how to determine submodels, switching points, and, subsequently, the mode sequence.

| Cluster no. | Submodel | $e_T + e_V$ | Term no. | Complexity index | Velocity range (m/s) | displacement range (m) |
|---|---|---|---|---|---|---|
| 9 | II | 2.9e-7 | 5 | 1.8 | [0.161 0.289] | [-0.058 0.074] |
| 15 | II | 3.63e-7 | 5 | 1.8 | [0.186 0.370] | [-0.059 0.074] |
| 25 | I | 1.15e-5 | 4 | 1.25 | [-0.026 0.029] | [0.079 0.101] |
| 26 | I | 1.27e-5 | 4 | 1.25 | [-0.019 0.038] | [0.079 0.101] |
| 35 | - | 0.074 | 6 | 1.5 | [0.051 0.134] | [0.057 0.099] |
| 35 (1) (85 data points) $z_2 \leq 0.104$ | I | 5.32e-4 | 4 | 1.25 | [0.051 0.104] | [0.058 0.099] |
| 35 (2) (15 data points) $z_2 > 0.104$ | II | 4.07e-4 | 5 | 1.8 | (0.104 0.134] | [0.057 0.096] |
| 45 | I | 1.13e-5 | 4 | 1.25 | [-0.029 0.023] | [0.079 0.101] |
| 40 | I | 8.24e-5 | 4 | 1.25 | [0.003 0.088] | 0.062 0.101 |

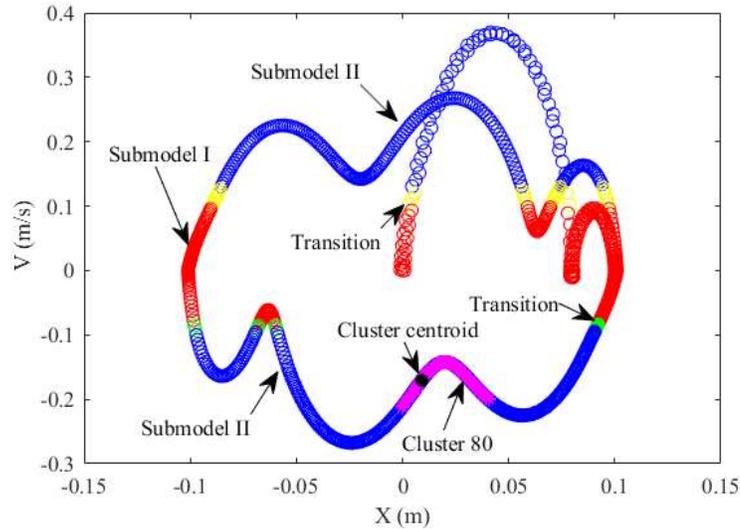

Fig. 12. The trajectory of the mass in the data-driven coordinate plane. The red circles belong to the submodel I and those in blue present the submodel II. The ones in green and yellow are clusters in which events occur.

# 8. Conclusion

An evolutionary symbolic sparse regression method was suggested for the system identification of multibody systems. The genetic programming was used to generate symbolic function expressions randomly and sparse regression approaches were used to obtain unknown coefficients of each function term in the equations of motion. A fitness measure was presented to promote parsimony in distilled equations and reduction in fit-to-data error. The capabilities of the developed algorithm were assessed considering three demonstrative examples. The robustness of the method was also illustrated by adding white Gaussian noise with zero mean to the time-series data and the case of incomplete dataset that requires time differentiation of noisy data was successfully investigated. The model did excel to estimate constant values of function arguments and discover rational functions that were not trivial to obtain. The hybrid dynamical systems were also considered, and a methodology was customized to determine the number of dynamic modes, respective submodels, and switching sequences successfully. The procedure demonstrated a good capability to identify not only the system parameters but also the governing motion equations of the system. It can be concluded that this technique can reduce the risk that the dictionary (a set of candidate functions) does not cover all functionality required to unravel hidden physical laws and the need for prior knowledge of the mechanism of interest. This is ongoing research and a future direction is to extend the procedure to experimentally acquired data. Moreover, this study plans to consider the possibility to discover the equations of motion with minimal coordinates from time-series data collected from multibody systems whose governing equations are written based on redundant coordinates as a future research work.

# Acknowledgments

The first author of this article would like to acknowledge the Marie Sktodowska-Curie Actions with Project No. 427452.

## Statements & Declarations

## Competing Interests

The authors have no relevant financial or non-financial interests to disclose.

## Conflict of interest

The authors declare that they have no conflict of interest concerning the publication of this manuscript.

## Data Availability

The data presented in this study are available on request from the corresponding author.